\documentclass[12pt]{article}
\usepackage{%
pstricks,%
pst-plot,%
epsf%
}                                  
\usepackage{cite}
\usepackage{array,dcolumn}      
\usepackage{axodraw}
\newcommand{\hs}{\hspace*{0.5cm}}
\newcommand{\be}{\begin{equation}}
\newcommand{\ee}{\end{equation}}
\newcommand{\bea}{\begin{eqnarray}}
\newcommand{\eea}{\end{eqnarray}}
\newcommand{\baa}{\begin{eqnarray*}}
\newcommand{\eaa}{\end{eqnarray*}}
\newcommand{\bet}{\begin{center} \begin{tabular}}
\newcommand{\ent}{\end{tabular} \end{center}}
\newcommand{\bary}{\begin{array}}
\newcommand{\eary}{\end{array}}
\newcommand{\bit}{\begin{itemize}}
\newcommand{\eit}{\end{itemize}}
\newcommand{\ra}{\rightarrow}

\newcommand{\crn}{\nonumber \\}

\newcommand{\al}{\alpha}
\newcommand{\ba}{\beta}
\newcommand{\ga}{\gamma}

\newcommand{\bb}{\begin{thebibliography}}
\newcommand{\eb}{\end{thebibliography}}
\newcommand{\ci}[1]{\cite{#1}}
\newcommand{\re}[1]{(\ref{#1})}
\newcommand{\bi}[1]{\bibitem{#1}}
\newcommand{\lab}[1]{\label{#1}}

\begin{document}
\begin{flushright}
BI-TP 99/44\\
hep-ph/9912304\\
November 1999
\end{flushright}
\begin{center}
{\large \bf Lepton flavour violating pion  decay
 $\pi^+ \ra \mu^-   \nu_\mu  e^+  e^+ $\\
and  the  $\mbox{SU}(3)_C\otimes \mbox{SU}(3)_L 
\otimes \mbox{U}(1)$  model }
\vspace*{1cm}

{\bf Hoang  Ngoc  Long}\footnote{On leave from 
 Institute of Physics, NCST, P. O. Box 429, Bo Ho, Hanoi 10000,
Vietnam}\\
\vspace*{0.3cm}

\medskip

{\it Fakult\"at f\"ur Physik, Universit\"at Bielefeld, 
D-33615 Bielefeld, Germany} \\ 
\vspace*{1cm}

{\bf Abstract}\\
\end{center}
\hs  In the framework of 
the minimal $\mbox{SU}(3)_C\otimes \mbox{SU}(3)_L 
\otimes \mbox{U}(1)_N$ model, the lepton-flavour-violating 
decay $\pi^+ \ra \mu^- \  \nu_\mu \ e^+ \ e^+ $ 
is calculated without directly invoking lepton mixing. 
The branching ratio for this rare
pion decay mode is found to be much smaller 
than the current experimental upper  limit. 
Dropping out  anomalous interactions,
this result coincides
with the previous calculation .

\vspace*{0.3cm}

\hs PACS number(s): 13.20.Cq, 13.20.Cz, 12.60.-i, 
12.10.Cn, 14.70.Pw.

\setcounter{footnote}{0}
\vspace*{0.5cm}


\hs At present, neutrinos are presumably massive
 and mixed 
as indicated in various experiments: 
 SuperKamiokande~\ci{suk} and  others~\ci{gri}.
This significant deviation from the  standard model (SM) calls
for its  extension. The models based on the
$\mbox{SU}(3)_C\otimes \mbox{SU}(3)_L \otimes \mbox{U}(1)_N$ (3 3 1) 
gauge group~\ci{ppf,rhn} are one of the most popular in 
such extensions beyond SM.
The SM assumes lepton-flavour-number conservation, and its 
observed violation would be a clear indication of new physics.
In the 3 3 1 models the lepton-flavour  number is not conserved,
and these models have motivated a variety of dedicated sensitive
searches for rare decay modes of muons and kaons and for
neutrino oscillations~\ci{fnos}.
It is known that the muon system is one of the best 
places to search for lepton flavour violation, compared with
the others. The ``wrong" muon decay
$\mu^- \ra e^- \nu_e {\bar \nu}_\mu$
 is widely used to put a lower bound on the singly-charged
bilepton mass ($M_Y \ge 230 $ GeV)~\ci{mu}.\par
\bigskip

\hs In this work we pay attention to the lepton-flavour-violating 
pion decay $\pi^+ \ra \mu^- \  \nu_\mu \ e^+ \ e^+ $. 
The upper limit in its branching ratio is given 
$ R  \le 1.6 \times 10^{-6} $ 
at  90 \% confidence level~\ci{pdg,bar}.
By suggesting the lepton mixing or  horizontal 
interactions, the above decay has been studied theoretically
in Ref.~\ci{theor}. However,
this decay may be described  by the 
minimal 3 3 1 model in  simple manner without directly
invoking lepton mixing. 

\hs To start, we firstly give some basic elements of the model
 (for more details see~\ci{dng}).
Three lepton components of each family  are in one triplet,
\[
f^{a}_L = \left( 
               \nu^a_L,\  l^a_L,\ (l^c)^a_L
                 \right)^T,
\]
where a = 1, 2, 3 is the family index. 
Under $SU(3)_L$, two of the three quark families  transform as 
antitriplet and one family transforms as triplet,
\[
Q_{iL} = \left( \begin{array}{c}
                d_{iL}\\-u_{iL}\\ D_{iL}\\ 
                \end{array}  \right), (i = 1,2), \ 
 Q_{3L} = \left( \begin{array}{c}
                 u_{3L}\\ d_{3L}\\ T_{L}
                \end{array}  \right).
\]
The right-handed quarks are singlets under  $SU(3)_L$. The exotic 
quarks $T$ and $D_i$ have an electric charge + 5/3 and 
- 4/3, respectively.

\hs There are five new gauge bosons: the $Z'$ and  
the charged bileptons with lepton number $L = \pm 2$, 
which are identified as follows:
$\sqrt{2}\ Y^-_\mu = W^4_\mu- iW^5_\mu ,\sqrt{2}$\ 
$ X^{--}_\mu =
W^6_\mu- iW^7_\mu $, and their 
couplings  to leptons are given by~\ci{framcal}:
\be
{\cal L}^{CC}_l = - \frac{g}{2\sqrt{2}}\left[
\bar{\nu}\gamma^\mu (1- \gamma_5) C\bar{l}^{T}Y^-_\mu 
 -  \bar{l}\gamma^\mu \gamma_5 C \bar{l}^T
X^{--}_\mu + \mbox{h.c.}\right].
\lab{ccl}
\ee

The interactions among the charged vector 
fields with quarks are
\bea
{\cal L}^{CC}_q &=&- \frac{g}{\sqrt{2}}[(\bar{u}_{3L}
\gamma^\mu d_{3L}+
\bar{u}_{iL}\gamma^\mu d_{iL})W^+_\mu +
(\bar{T}_{L}\gamma^\mu d_{3L}+\bar{u}_{iL}
\gamma^\mu D_{iL})X^{++}_\mu 
\nonumber \\
                 & & + (\bar{u}_{3L}\gamma^\mu T_{L}-
\bar{D}_{iL}\gamma^\mu 
d_{iL})Y^{-}_\mu + \mbox{h.c.}].
\lab{ccq}
\eea
It is to be noted that the vector currents 
coupled to $X^{--}$,
$X^{++}$ vanish due to Fermi statistics, and
the exotic quarks interact with ordinary ones only via the 
bileptons and non-SM Higgs bosons.

\hs The current experimental lower bound on the exotic quark mass is 
200 GeV~\ci{jm}, while  the lower bound on the
bilepton mass is in the range of 300 GeV.

\hs To deal with the above process we also need the
coupling constants of the bileptons $X,\ Y$ to the SM 
weak-vector boson $W$. In the  notation of Refs.\ci{grace} 
it is: $CWXY = \frac{g}{\sqrt{2}}$.

\hs Now we start with the decay
\be
\pi^+(K) \ra \mu^-(p) \  \nu_\mu(q) \ e^+(k_1) \ e^+(k_2), 
\lab{qt}
\ee
where the  letters in parentheses stand for the momenta of the particles.
We assume that the Higgs bosons responsible for lepton-flavour-violating 
interactions as well as the exotic quarks
 are much heavier than the standard model $W$ boson.
Hence the contributions from the exotic quarks and
non SM Higgs bosons are suppressed.
With new gauge bosons carrying lepton-number $L = 2$,
the process \re{qt} can be described simply by the 
Feynman diagram depicted in Fig. 1.

\hs For small momentum-transfer ($ q^2 << m_W^2, M_X^2, M_Y^2$), as is 
the case here, the matrix element for this process is found to be
\bea
{\cal M}_{fi} &=& 2 \frac{G^2_F f_\pi m_W^2}{M_X^2 \ M_Y^2}
\left[ -(P+K)_\ba K_\ga + (K+L)_\ga K_\ba + (-L+P).K\  g_{\ba \ga}
\right]\crn
& & \times {\bar u}_{\nu_\mu}(q) \ga^\ba (1-\ga_5)C 
{\bar u}^T_\mu(p). v^T_e(k_1)C\ga^\ga\ga_5v_e(k_2),
\lab{mat}
\eea
where the following combinations of four vectors are introduced
\be
P= k_1 + k_2,\ Q= k_1 - k_2,\ L= p+q, \ N=p-q, \ K= P + L.
\ee
\vspace*{0.9cm}

\begin{center}
\begin{picture}(260,50)(-5,0)
\Photon(11,10)(52,10){2}{4}
\Photon(52,10)(80,50){2}{4}
\Photon(52,10)(80,-30){2}{4}
\ArrowLine(80,50)(115,33)
\ArrowLine(80,50)(115,67)
\ArrowLine(115,-47)(80,-30)
\ArrowLine(115,-13)(80,-30)
\DashLine(-20,11)(11,11){2}
\DashLine(-20,10)(11,10){2}
\DashLine(-20,9)(11,9){2}
\DashLine(-20,8)(11,8){2}
\Vertex(11,10){1.8}
\Text(-5,20)[]{$\pi^+$(K)}
\Text(25,0)[]{$W^+$}
\Text(55,30)[]{$Y^-$}
\Text(55,-20)[]{$X^{++}$}
\Text(132,33)[]{$\nu_\mu(q)$}
\Text(132,67)[]{$\mu^-(p)$}
\Text(134,-47)[]{$e^+(k_2)$}
\Text(134,-13)[]{$e^+(k_1)$}
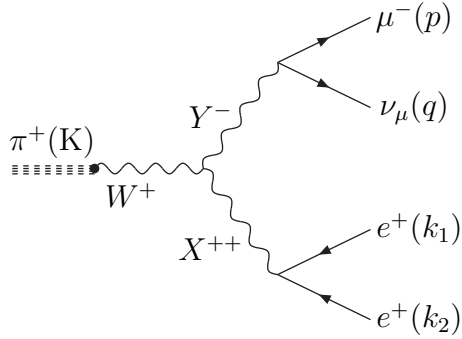
\Text(90,-80)[]{ Figure 1: Feynman diagram for the decay
$\pi^+(K) \ra \mu^-(p) \  \nu_\mu(q) \ e^+(k_1) \ e^+(k_2)  $}
\Text(90,-90)[]{ in the 3 3 1 model }
\end{picture}
\end{center}

\vspace*{3.5cm}

\hs The squared matrix element is given by
\bea
|{\cal M}_{fi}|^2 & =& 128\frac{G_F^4 f_\pi^2 m_W^4}{M_X^4 M_Y^4}
C_{\ba \ga} C_{\ba' \ga'}\crn
& & \times \left[ p^\ba q^{\ba'} +  p^{\ba'} q^{\ba} - 
g^{\ba \ba'} (p.q) + i\varepsilon^{\ba \ba' m n}p_m q_n\right]\crn
& & \times \left[ k_1^{\ga} k_2^{\ga'} +  k_1^{\ga'} k_2^{\ga}
- g^{\ga \ga'}(k_1.k_2 -m_e^2)\right],
\lab{bfmt}
\eea
where the notation $C_{\ba \ga}\equiv
\left[ -(P+K)_\ba K_\ga + (K+L)_\ga K_\ba + (-L+P).K\  g_{\ba \ga}
\right]$ is used.

\hs In order to describe the kinematics of the decay, we introduce the 
following vectors: Let ${\vec v}$ be a unit vector along 
the direction of flight of the dipositron in the $\pi^+$ rest 
system $(\Sigma_\pi)$, ${\vec a}$ be  a unit vector along the 
projection of the three-momentum of the $e^+$ in  the $e^+\  e^+$ 
center-of-mass system $(\Sigma_{2e})$  perpendicular to  ${\vec v}$, 
and  ${\vec b}$ be a unit vector  along the projection 
of the three-momentum of the $\mu^-$ in the 
$\mu^- \ \nu_\mu$ center-of-mass system $(\Sigma_{\mu \nu})$  
perpendicular to  $-{\vec v}$.
Then the kinematics of this decay is similar to the one given in~\ci{kb},
which consists of five variables: $s_e  \equiv P^2$,  
$s_\mu  \equiv L^2$, and three angles:(i)
 $\theta_e$, the angle of the $e^+$ in 
$\Sigma_{2e}$ with respect to the dipositron 
line of flight in $\Sigma_\pi$,
(ii) $\theta_\mu$, the angle of the $\mu^-$ in  $\Sigma_{\mu \nu}$ 
with respect to the $\mu \nu_\mu$ line of flight in $\Sigma_\pi$, 
and  (iii) $\phi$, the angle between the plane formed by the positrons in
$\Sigma_\pi$  and the corresponding plane formed by the $\mu^-, \nu_\mu$.
The angles $\theta_e,\theta_\mu$ and $\phi$ are shown in Fig. 2.
\vspace*{0.1cm}

\begin{center}

\begin{picture}(260,50)(-5,0)
\Line(50,0)(65,-50)
\Line(50,0)(35,-50)
\Line(-65,-50)(35,-50)
\Line(-65,-50)(-35,50)
\Line(65,-50)(165,-50)
\Line(133,54)(165,-50)
\Line(50,0)(32,54)
\Line(-35,50)(35,50)
\DashLine(50,0)(65,50){2}
\DashLine(35,50)(65,50){2}
\DashLine(-46,0)(40,0){2}
\DashLine(65,0)(140,0){2}
\DashLine(-15,-50)(15,50){2}
\DashLine(115,-50)(85,50){2}
\Line(133,54)(32,54)
\ArrowLine(0,0)(-7,25)
\Text(-7,30)[]{$e^+$}
\ArrowLine(0,0)(7,-25)
\Text(7,-30)[]{$e^+$}
\ArrowLine(0,0)(6,20)
\Text(13,20)[]{${\vec a}$}
\ArrowLine(100,0)(125,20)
\Text(125,30)[]{$\mu^-$}
\ArrowLine(100,0)(94,20)
\Text(87,20)[]{${\vec b}$}
\ArrowLine(100,0)(85,-12)
\Text(85,-20)[]{$\nu_\mu$}
\ArrowLine(50,0)(35,0)
\Text(30,5)[]{${\vec v}$}
\Text(60,0)[]{${\pi^+}$}
\ArrowArcn(0,0)(15,180,100)
\Text(-22,7)[]{${\theta_e}$}
\ArrowArc(100,0)(15,0,45)
\Text(127,7)[]{${\theta_\mu}$}
\ArrowArc(50,0)(20,75,105)
\Text(50,30)[]{${\phi}$}
\Text(-38,-45)[]{e-e plane}
\Text(130,-45)[]{$\mu - \nu$ plane}
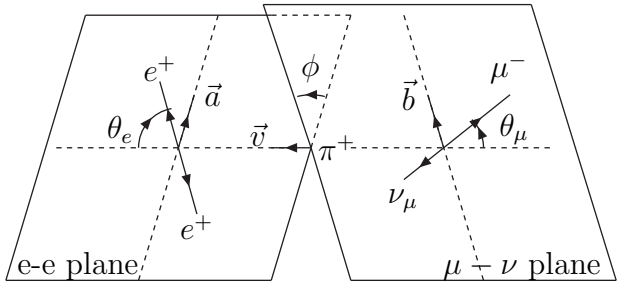
\Text(90,-80)[]{ Figure 2: Illustration of the angles $\theta_e,\
\theta_\mu$ and $\phi$}
\end{picture}
\end{center}
\vspace*{3cm}

\hs Then the decay width for the pion decay \re{qt} is written as 
\be
d\Gamma =\left(\frac{1}{2}\right) \frac{1}{2^{14}\pi^6 m_\pi^3}
\sum_{spins}|{\cal M}_{fi}|^2
  \left(1-z_\mu\right)\sigma_e X d s_e d s_\mu d(\cos_e)
d(\cos_\mu) d\phi.
\lab{td}
\ee
In \re{td}, $\left( \frac{1}{2}\right)$  is the statistical factor 
indicating that two (identical) positrons in the final state~\ci{cl}.
With the above  definitions we have the following scalar products
\bea
 Q^2& = & 4 m_e^2 - s_e,\ N^2 = 2 m^2_\mu - s_\mu, \  
K^2 = m_\pi^2, \ L.N = m^2_\mu,  \crn
P.L &=& \frac{1}{2}(m^2_e -s_e - s_\mu), \ 
P.N = z_\mu P.L + (1 - z_\mu) X \cos \theta_\mu, \ 
Q.L = \sigma_e X \cos \theta_e,\\
Q.N & =& z_\mu Q.L + \sigma_e (1-z_\mu) P.L \cos \theta_e \cos 
\theta_\mu - (s_e s_\mu)^{1/2} \sin \theta_e \sin 
\theta_\mu \cos \phi, \crn
d& \equiv&  \varepsilon^{\mu \nu \al \ba} L_\mu N_\nu P_\al Q_\ba 
 = - (s_e s_\mu)^{1/2} \sigma_e (1-z_\mu) X \sin \theta_e
\sin \theta_\mu \sin \phi,\nonumber
\eea
where
\[z_\mu \equiv 
\frac{m^2_\mu}{s_\mu},\  \sigma_e\equiv 
\left(1 - \frac{4 m^2_e}{s_e}\right)^{1/2},\
 X\equiv ((P.L)^2 - s_e s_\mu)^{1/2},\]
and $m_e, m_\mu, m_\pi $ stand for masses of the electron,
the muon and the pion, respectively.\\
The range of the variables is
\bea
4 m_e^2& \leq& s_e \leq (m_\pi - m_\mu)^2,\crn
m_\mu^2& \leq& s_\mu \leq (m_\pi -\sqrt{s_e})^2,\\
0&\leq& \theta_e, \theta_\mu \leq \pi, \ 0 \leq \phi \leq 
2 \pi.\nonumber
\lab{gh}
\eea

\hs It is to be noted that an imaginary part of $|{\cal M}_{fi}|^2 $
connected with pseudotensor $d$
is {\it linear} in $\sin \phi$, i.e. no such a term like $Q.N d$, 
hence it will be removed after integration over the angle $\phi$.
In resulting we get the decay width being a real number, 
as it has to be.

\hs The integrations over the angles can be carried out 
analytically by using Mathematica. The numerical 
integrations over the effective masses squared  $s_e$ 
and $s_\mu$ are carried 
out by employing the Monte Carlo routine VEGAS~\ci{ve}.
After changing  to dimensionless
parameters $x_e=\frac{s_e}{m^2_\pi}, \ 
y_\mu=\frac{s_\mu}{m^2_\pi}$,  we get the decay width
\be
\Gamma (\pi^+ \ra \mu^- \  \nu_\mu \ e^+ \ e^+) = 
\frac{  G_F^4 f^2_\pi m_W^4 m_\pi^{11} N
 }{256 \pi^6 M_X^4 M_Y^4},
\lab{srmi}
\ee
where $N$ is numerically evaluated,  
$N =  6.17 \times 10^{-6}$.
\hs We recall  that the main (99.987 \%) decay mode of the $\pi^+$ is 
well-known
\be
\Gamma(\pi^+ \ra \mu^+ \  \nu_\mu) = \
\frac{G_F^2 f^2_\pi m_\mu^2}{8 \pi  m_\pi^3}(m_\pi^2 -m_\mu^2)^2
\simeq 2.63 \times 10^{-17}\  {\mbox GeV}.
\lab{smr}
\ee
From \re{srmi} and \re{smr} we get the branching ratio
\bea
R_\pi& = &\frac{\Gamma (\pi^+ \ra \mu^- \  \nu_\mu \ e^+ 
\ e^+)}{\Gamma(\pi^+ \ra \mu^+ \  \nu_\mu)} = 
\frac{ 6.17 \times 10^{-6}\  G_F^2  m_W^4 m_\pi^{14}}{32
\pi^5  M_X^4 M_Y^4 m_\mu^2 (m_\pi^2 -m_\mu^2)^2}\crn
&\simeq& 4.97 \times 10^{-18}\frac{1}{M_X^4 [GeV] M_Y^4 [GeV]}. 
\lab{ra}
\eea
Putting $M_X \simeq M_Y \simeq 120 $ GeV as a lower limit
 obtained from the LEP data analysis~\ci{fram}, 
we get $R_\pi \sim 2.3 \times 10^{-34}$.
This number is much smaller than the current experimental upper
limit, but it coincides with the previous theoretical
evaluation without anomalous interactions included~\ci{bar}.
It rises a question about the mechanism for large
lepton-flavour-violating pion decay mode.
However, it is worth  mentioning that the experimental
data on $R_\pi$  decrease with time, for example the 1988 data
were $R_\pi \leq 8 \times 10^{-6}$, while the 1998 data are
 $R_\pi \leq 1.6 \times 10^{-6}$.
 We suggest that by adding contributions from 
diagrams with the exotic quarks and Higgs bosons the situation will
be modified but not improved too much.

\hs Our calculation can be analogously applied for the 
lepton-flavour-violating kaon decay 
 $K^+ \ra \mu^-   \nu_\mu  e^+  e^+ $, which has an experimental
branching ratio of $R_K \leq 2.0 \times 10^{-8}$. However, the main 
decay mode $K^+ \ra \mu^+ \  \nu_\mu$ has only a branching ratio of 
69.51 \%, instead of 99.987\% in the $\pi^+$ case considered here .

\hs In summary, we have considered the lepton-flavour-violating 
pion decay without directly invoking lepton mixing.
Our result is by  twenty eight  orders smaller than the current 
experimental upper
limit. This conclusion should not be modified too much
by including contributions from the exotic quarks and Higgs bosons.
Hence, the mechanism for  large lepton-flavour-violating pion
decay mode is a mystery.

{\it Acknowledgments}

\hs This work was supported by the DAAD grant.
The author thanks Prof. D. Schildknecht for  
stimulating remarks and S. Dittmaier for consultation 
on the Monte Carlo routine VEGAS.
This work was initiated when the author was at LAPTH, 
Annecy, France. He thanks G. Belanger  and F. Boudjema 
for consultation on the decay kinematics  and a 
numerical calculation. 
Thanks are due to P. Aurenche and Theory Group at 
LAPP and  Bielefeld University  for kind
hospitality.

\end{document}